\RequirePackage{fix-cm}
\documentclass[smallextended]{svjour3} 
\smartqed 
\usepackage{graphicx}
\usepackage{caption}
\usepackage{nicefrac}
\usepackage{amsfonts}
\usepackage[T1]{fontenc}
\usepackage{xspace}
\usepackage{flushend}
\usepackage{ifthen}
\usepackage{color}
\usepackage{url}
\usepackage{multirow}
\usepackage{listliketab}
\usepackage{amssymb}
\usepackage[linesnumbered,lined,boxed,commentsnumbered]{algorithm2e}
\usepackage{algorithmic}
\usepackage{booktabs}
\usepackage{comment}
\usepackage{amsmath}
\usepackage{bbding}
\usepackage{listings}
\usepackage{float}
\usepackage[misc]{ifsym}
\usepackage{tabularx}
\usepackage{rotating}
\usepackage{xcolor}
\usepackage{natbib}
\usepackage{tcolorbox}
\usepackage{ulem}
\usepackage{censor}
\usepackage{subcaption}
\usepackage{pgfplots}
\pgfplotsset{compat=1.14}
\usepackage{pgfplotstable}
\usepackage[colorinlistoftodos]{todonotes}
\def\mybar#1{%%
  #1s & {\color{grey}\rule{#1cm}{8pt}}}
\definecolor{Fork}{HTML}{99CCFF}
\definecolor{Non-Fork}{HTML}{FFCCCB}

\def\mybar#1{%%
  {\color{gray}\rule{#1cm}{8pt}}}

\usepackage{tikz}

\colorlet{punct}{red!60!black}
\definecolor{background}{HTML}{EEEEEE}
\definecolor{delim}{RGB}{20,105,176}
\colorlet{numb}{magenta!60!black}

\lstdefinelanguage{json}{
    basicstyle=\small,
    numbers=left,
    numberstyle=\scriptsize,
    stepnumber=1,
    numbersep=8pt,
    showstringspaces=false,
    breaklines=true,
    frame=lines,
    backgroundcolor=\color{background},
    literate=
     *{0}{{{\color{numb}0}}}{1}
      {1}{{{\color{numb}1}}}{1}
      {2}{{{\color{numb}2}}}{1}
      {3}{{{\color{numb}3}}}{1}
      {4}{{{\color{numb}4}}}{1}
      {5}{{{\color{numb}5}}}{1}
      {6}{{{\color{numb}6}}}{1}
      {7}{{{\color{numb}7}}}{1}
      {8}{{{\color{numb}8}}}{1}
      {9}{{{\color{numb}9}}}{1}
      {:}{{{\color{punct}{:}}}}{1}
      {,}{{{\color{punct}{,}}}}{1}
      {\{}{{{\color{delim}{\{}}}}{1}
      {\}}{{{\color{delim}{\}}}}}{1}
      {[}{{{\color{delim}{[}}}}{1}
      {]}{{{\color{delim}{]}}}}{1},
}

\begin{document}

\title{Does the First Response Matter for Future Contributions? A Study of First Contributions}

\author{Noppadol Assavakamhaenghan \Letter\and
        Supatsara Wattanakriengkrai\and
        Naomichi Shimada\and
        Raula Gaikovina Kula\and
        Takashi Ishio\and
        Kenichi Matsumoto%etc.
}

\institute{
    \Letter~Corresponding author - Noppadol Assavakamhaenghan
    \at Nara Institute of Science and Technology, Japan\\
    \email{\{noppadol.assavakamhaenghan.mt8\}@is.naist.jp}
     \and
    \email{\{wattanakri.supatsara.ws3, shimada.naomichi.sm3, raula-k, ishio, matumoto\}@is.naist.jp}
}

\date{Received: date / Accepted: date}

\maketitle

\begin{abstract}
Open Source Software (OSS) projects rely on a continuous stream of new contributors for their livelihood. 
Recent studies reported that new contributors experience many barriers in their first contribution, with the social barrier being critical. 
Although a number of studies investigated the social barriers to new contributors, we hypothesize that negative first responses may cause an unpleasant feeling, and subsequently lead to the discontinuity of any future contribution.
We execute protocols of a registered report to analyze 2,765,917 first contributions as Pull Requests (PRs) with 642,841 first responses.
We characterize most first response as being positive, but less responsive, and exhibiting sentiments of fear, joy and love.
Results also indicate that negative first responses have the literal intention to arouse emotions of being either constructive (50.71\%) or criticizing (37.68\%) in nature.
Running different machine learning models, we find that predicting future interactions is low (F1 score of 0.6171), but relatively better than baselines. 
Furthermore, an analysis of these models show that interactions are positively correlated with a future contribution, with other dimensions (i.e., project, contributor, contribution) having a large effect.
\keywords{Sentiment and Toxicity, First Response, Newcomer}
\end{abstract}

\section{Introduction and Motivation}
\label{sec:introduction}
GitHub is a well-known repository hosting platform for Open Source Software (OSS) projects. 
In 2021, GitHub reported over 60 million repositories created in recent years and over 56 million contributors; a massive number of new contributors strive to present their skills to the world's largest OSS community through project contributions. 
GitHub strives on social coding, which is the collaboration of both new contributors and the current community of contributors \citep{Choi2010}.
OSS projects rely on a continuous stream of new contributors for their livelihood.
% \citep{Steinmacher2013}
Recent studies provide evidence that new contributors have encountered various barriers ranging from difficulty in finding assistance to receiving negative responses from other contributors. 
% \citep{Steinmacher2015}. 
% As a consequence, the rate of new contributors has been on the increase \citep{Steinmacher2014}.
% For example, \cite{Steinmacher2013} reports that less than 18\% of new contributors keep contribution.
% This loss possibly impacts OSS projects' livelihood and sustainability \citep{Valiev_2018}.
% % However, the social barrier has been found to be one of the critical barriers that hinder the new contributors' first contribution \cite{Steinmacher2015}.
% \cite{6227164} revealed the probability for a new contributor to becoming a long-term contributor is associated with his/her willingness and environment, who are socially welcomed and given constructive criticism. 
%  \cite{Gousios_2016} reported that responsiveness is the most reported challenge that new contributors experience.
% \todoMap{R4.1}{tor}

\begin{figure*}[!]
\centering
\includegraphics[width=\linewidth]{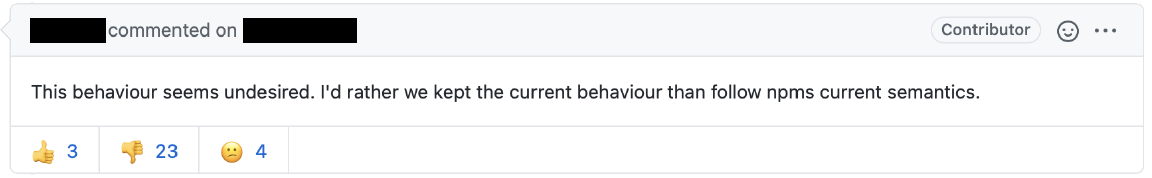}
\caption{A screenshot example of a negative response to a new contributor, which may deter them interacting in the future. }
\label{fig:ex_comment}
\end{figure*}

% \raula{move Fig 1 to page 2\checkmark}
Figure \ref{fig:ex_comment} shows a screenshot of a negative response to a first contribution in GitHub. 
In terms of GitHub projects, we consider any pull request (PR) as a contribution, with comments exchanged during the review of PR as interactions between contributors and reviewers. 
Hence, a first response is the first comment to PR submitted by a newcomer to that repository.
In this case:
\begin{quote}
    \textit{``This behavior seems undesired. I'd rather we kept the current behavior than follow npms current semantics."}
\end{quote}
We hypothesize that such negative messages may cause an unpleasant feeling and subsequently lead to the discontinuity of their contributions (i.e., such as further interactions within the same PR or future contributions), especially for a novice contribution that can be easily demotivated.
% Although a number of studies investigated the social barriers to new contributors, to the best of our knowledge, the relationship between the first response to the first contributions and their future contributions has not been studied comprehensively. 

In this study, we collect and analyze 72,635 active projects in GHTorrent dataset \citep{ghtorrent}. From these projects, we then extracted 2,765,917 PR and 642,841 first response contained in the PR.
The execution protocol included both qualitative and quantitative approaches that are driven by three research questions:

\begin{itemize}
    \item \textbf{
RQ1: {What are the characteristics of the first responses toward first contributions?}}
\textit{Results - }
From statistical testing of sentiment, responsiveness, toxicity and emotions of first responses, we find that first contributions are more likely to have positive sentiments (\textit{H1.1}), but are less responsive (\textit{H1.2}) compared to non-first contributions. 
Furthermore, we show that toxic responses are not bias to the first contribution (\textit{H1.3}), however, developers did expressed certain sentiments (i.e., fear, joy and love) specific to first contribution responses (\textit{H1.4}).
    \item \textbf{
RQ2: {What is the relationship between receiving a positive response and further interactions in the same PR for a first-time contributor?}} \textit{Results - }
An investigation into positive first responses reveals that a positive (\textit{H2.1}) and responsive response (\textit{H2.2}) is not in relation to an increase in the likelihood for future interaction.
In terms of negative first responses, qualitatively, we find that most comments are either suggestions (51.97\%), comments (32.28\%) and questions (30.71\%), and had attributes of being either constructive (51.97\%) or criticizing (37.53\%).
    
\item \textbf{RQ3: {What is the relationship between first-time contributors' interactions and their future contributions?}} \textit{Results - }
We analyze four different dimensions (interactions, contributor, project, contribution) to find that contributor interactions attributes have a positive correlation to having a future contribution.
Furthermore, we identified several important predictive features related to project (\#PR, \#commit, language),  contributor (GitHub account), and  contribution (size).
    
\end{itemize}

While executing the protocols, we encountered unavoidable deviations.
Deviations are related to our data (Section \ref{sec:Data_Preparation}), metrics, analysis (Section \ref{sec:sanitycheck}, \ref{sec:RQ1}, and \ref{sec:RQ2}), and methodology (Section \ref{sec:sanitycheck} and \ref{sec:RQ2}) 
A full listing of the deviations is contained in our replication package \footnote{\url{https://github.com/NAIST-SE/FirstResponsePR/blob/main/DEVIATIONS.md}}.
The remainder of the paper is organized as follows.
Section \ref{sec:Data_Preparation} provides the data preparation processes for our datasets. Section \ref{sec:Finding} provides hypothesis, approach, and findings in each RQ. Section \ref{sec:implication} discusses implications of our work. Section \ref{sec:threats} exposes potential threats to validity. Section \ref{sec:Related_Work} reviews previous studies related to our study. Section \ref{sec:conclusion} concludes the paper.
We make the datasets available at: \url{https://github.com/NAIST-SE/FirstResponsePR}

\section{Data Preparation}
\label{sec:Data_Preparation}
In this section, we describe our tools and the data preparation process used to answer our research questions. 

\subsection{Textual Analysis Tools}
The state-of-the-art text analysis tools used for extracting sentiment, toxicity, and emotion of the first responses in our study are listed below:
\begin{itemize}
    \item \textbf{\textit{Sentiment analysis.}} The SentiStrength-SE tool by \citet{sentiment_tool} is a sentiment analysis tool that utilize domain dictionary and heuristics for software engineering text, which outperforms existing domain-independent tools/toolkits (SentiStrength, NLTK, and Stanford NLP)  for the software engineering. Input is the first response and the output is sentiment score ranging from -5  (very negative) to 5 (very positive).
    \item \textbf{\textit{Toxicity detection.}} The toxicity detection tool by \citet{toxicity} combines the output of the Perspective API\footnote{\url{https://www.perspectiveapi.com/}} with the output from a customized version of the Stanford’s Politeness detector tool trained on software engineering text. Input is the first response and the output is binary value.
    \item \textbf{\textit{Emotion detection.}} The EmoTxt tool by \citet{emotion} utilizes Support Vector Machine. It is trained on two gold standard datasets derived from 4,800 posts from Stack Overflow and 4,000 comments posted by software developers on Jira. Input is the first response and the output is binary value that shows whether the text contains of "Joy", "Anger", "Sadness", "Love", "Surprise", and "Fear."
\end{itemize}

\subsection{Data Collection}

\begin{figure*}[!t]
\centering
\includegraphics[width=\linewidth]{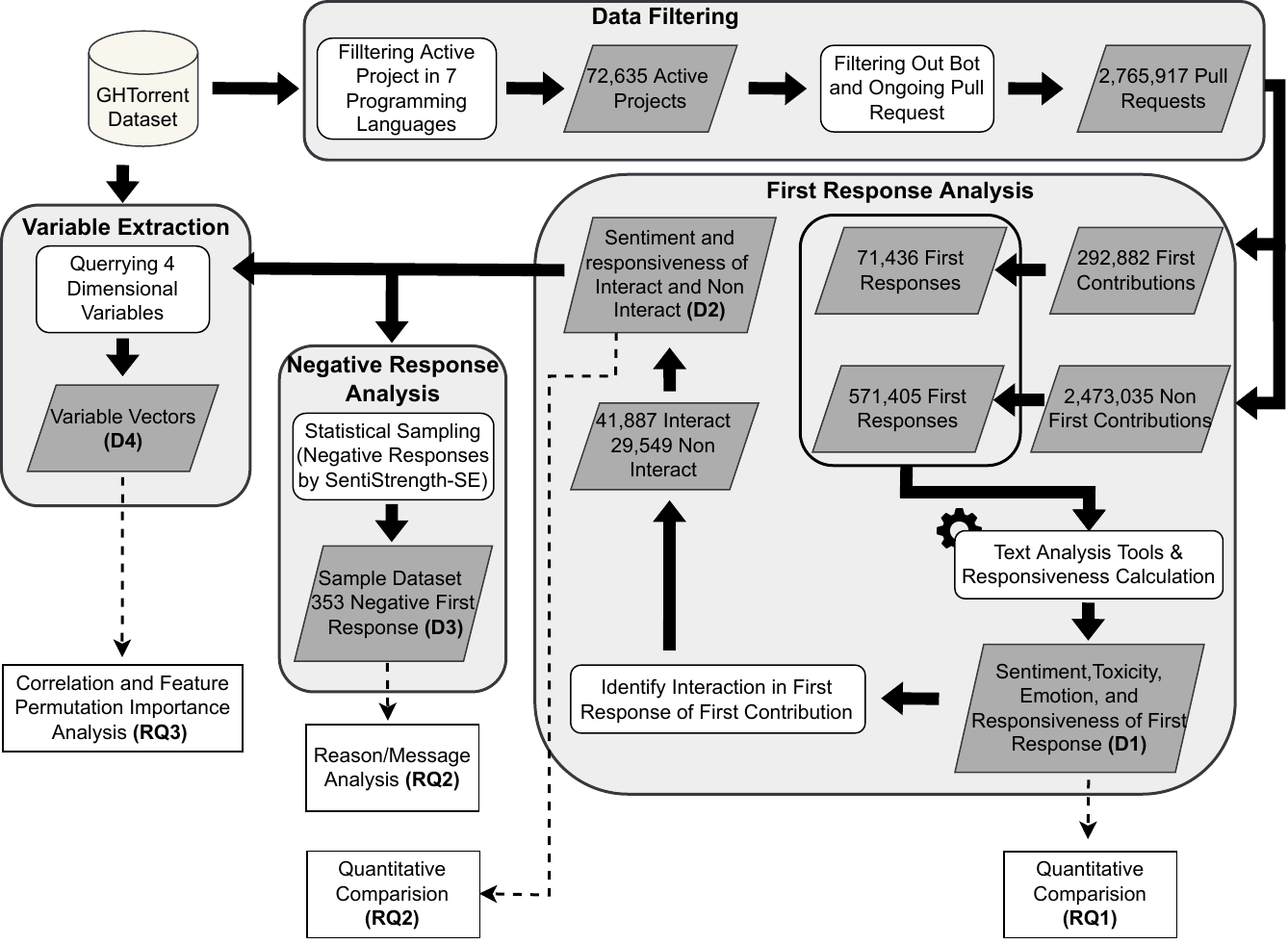}
\caption{Data Preparation Overview}
\label{fig:overview}
\end{figure*}

Figure \ref{fig:overview} presents an overview of the data preparation process.
The starting point of our data preparation process is selecting the data source that contain PRs information.
As a deviation to the registered report, we instead use GHTorrent\footnote{updated as to 2021/03/06} as our data source.
Due to its large SQL database size (over 300 GB in total), we use Pyspark\footnote{\url{https://spark.apache.org/}} to reduce query time.
A summary of the collected dataset is shown in Table \ref{tab:dataset_statistic}. 

\begin{table}[]
\caption{Summary of Data Collected}
\label{tab:dataset_statistic}
\centering
\begin{tabular}{@{}lr@{}}
\toprule
\multicolumn{2}{c}{Projects}                                                                            \\ \midrule
\multicolumn{1}{l|}{\# Projects from GHTorrent}                                   & 30,434,411                        \\
\multicolumn{1}{l|}{\# Projects after Filtering}             & 72,635                   \\ \midrule
\multicolumn{2}{c}{Contributions}                                                                       \\ \midrule
\multicolumn{1}{l|}{\# PRs after Filtering}                              & 13,676,709               \\
\multicolumn{1}{l|}{\# Closed PRs}                                   & 2,765,917                \\
\multicolumn{1}{l|}{\# First Contributions}                                    & 10.59\% (from 2,765,917) \\ \bottomrule
\end{tabular}
\end{table}

\begin{table}[]
\caption{Summary for each dataset (D1, D2, D3, and D4)}
\label{tab:dataset_create}
\centering
\begin{tabular}{@{}lr@{}}
\toprule
\multicolumn{2}{c}{First Responses}                                                                       \\ \midrule
\multicolumn{1}{l|}{\# Non-First Contributions(D1)}      & 571,405                  \\
\multicolumn{1}{l|}{\# First Contributions(D2)}          & 71,436                   \\
\multicolumn{1}{l|}{\# with Interaction (D2)}                  & 58.66\% (from 71,436)    \\\midrule
\multicolumn{1}{l|}{\# Negative First Responses}                                & 4,353                    \\
\multicolumn{1}{l|}{\# Negative First Response (D3)} & 353                      \\ \midrule
\multicolumn{2}{c}{Contributions}                                                                       \\ \midrule
\multicolumn{1}{l|}{\# First Contributions {(D4)}}     & 50.45\% (from 71,436)    \\ \bottomrule
\end{tabular}
\end{table}
\textbf{Data Filtering.}
We apply two filters. The first filter is to ensure active projects. Similar to related work \citep{Hata_dataset}, we filter projects that fit the criteria of: (i) having more than 500 commits in their entire history, and (ii) having at least 100 commits in each of the most active two years.
From the first filter, we are left with 72,635 projects with 13,676,709 PRs. 

The second quality filter is to select only PRs that are "closed" and are not created by bots. To do this systematically, we use state-of-the-art tools \citep{botdetection}, removing PRs created by 3,039 bots.
Finally, we are left with 2,765,917 PRs. 
From our dataset, we then created different datasets that are used to answer specific research questions. Details of all datasets all statistics are shown in Table \ref{tab:dataset_create}. 

\paragraph{\textbf{First Response/Interaction Analysis (Datasets D1 and D2).}}

Dataset D1 contains textual and responsiveness data between first contributions and non-first contributions. It is extracted in three steps. The first step is to distinguish a first contribution using the SQL query\footnote{\label{experiment_notebook}\url{https://github.com/NAIST-SE/FirstResponsePR/blob/main/Experiment.ipynb}} in Section "Extract first contribution and non-first contribution", leaving us with 292,882 first contributions and 2,473,035 non-first contributions. 
For the second step, we identify and extract the first response. Using PRs that were extracted from step one, we then query our dataset using the SQL query\footnotemark[\value{footnote}] in Section "Extract comment in PR of inspect project", obtaining 71,436 first reviewing responses from first and 571,405 from non-first contributions. To calculate responsiveness of the first responses, we measure the duration between creation of PR and the time when the first comment was posted. 

Dataset D2 contains the interactions of first contributions. As a subset of D1, D2 first extracts sentiment and responsiveness.
We also distinguish between those first contributions that included an interaction, defining an interaction as to when a contribution contains at least on reply to the first response. 
Refer to the script in our repository\footnotemark[\value{footnote}] in Section "Extract Interact after first response of first time contributor."
In the end, we obtain 41,887 first response with interaction and 29,549 without interaction.  

\paragraph{\textbf{Negative Response Analysis (Dataset D3)}}
Dataset D3 is used to answer RQ2. 
To create the dataset, we first extract the negative first responses in first contributions, where sentiment score is less than zero.
This results in 4,453 of negative first responses. 
We then take a qualitative sample of 353 responses, using a 95\% confidence level and 5\% confidence interval\footnote{\label{surveysystem}\url{https://www.surveysystem.com/sscalc.htm}}.

\paragraph{\textbf{Variable Extraction (Dataset D4)}}
Dataset D4 is used to answer RQ3. 
D4 contains both the first response from D2 and four different dimensions of variables (interactions, contributor, project, contribution) from the registered report\citep{Assavakamhaenghan:msr2021}. 
There are 36,043 of positive class (first-time contributors contribute again after their first contribution) and 35,393 of negative class sample in the dataset.
Hence, D4 contains textual analysis and the four dimension variables (cf. Section 3.3).

\begin{table}[]
\caption{Evalution of the Sanity Check}
\label{tab:acc_kappa_sanity}
\centering
\begin{tabular}{@{}llrrrrr@{}}
\toprule
\multicolumn{2}{c}{Textual Analysis} & \multicolumn{1}{c}{Accuracy} & \multicolumn{1}{c}{Precision} & \multicolumn{1}{c}{Recall} & \multicolumn{1}{c}{F1} & \multicolumn{1}{c}{Distribution} \\ \midrule
Sentiment & \multicolumn{1}{l|}{} & 0.86 & - & - & - & (25, 333, 26) \\
Toxicity & \multicolumn{1}{l|}{} & 0.96 & 0.00 & 0.00 & 0.00 & (2, 382) \\
\multirow{6}{*}{Emotion} & \multicolumn{1}{l|}{Joy} & 0.88 & 0.20 & 0.16 & 0.18 & (25, 359) \\
 & \multicolumn{1}{l|}{Anger} & 0.92 & 0.50 & 0.07 & 0.12 & (4, 380) \\
 & \multicolumn{1}{l|}{Sadness} & 0.94 & 0.37 & 0.41 & 0.39 & (19, 365) \\
 & \multicolumn{1}{l|}{Love} & 0.95 & 0.43 & 0.18 & 0.25 & (7, 377) \\
 & \multicolumn{1}{l|}{Surprise} & 0.88 & 0.53 & 0.61 & 0.56 & (59, 325) \\
 & \multicolumn{1}{l|}{Fear} & 0.90 & 0.41 & 0.45 & 0.43 & (34, 350) \\ \bottomrule
\end{tabular}
\end{table}

\subsection{Sanity Check}
\label{sec:sanitycheck}
As outlined in the registered report, we perform a manual confirmation of the tool performance, as well as validation of the contributor identity. We draw a representative sample. The sample size is calculated with confidence level of 95\% and confidence interval of 5\%\footnotemark[\value{footnote}], resulting in 384 samples.   

We follow procedures for a Kappa agreement\citep{KappaAgreement} to validate our tool outputs.
The kappa agreement score is interpreted as follows: No agreement (N) $k < 0$, Slight agreement (Sl) $0.0 \leq k \leq 0.2$, Fair  agreement (F) $0.2 < k \leq 0.4$, Moderate agreement (M) $ 0.4 < k \leq 0.6$, Substantial agreement (Su) $ 0.6 < k \leq 0.8$, and Almost perfect agreement (AP) $ 0.8 < k \leq 1.0$. 
For the initial iteration, the first four authors individually labeled 30 responses, indicating whether or not they agreed with the result.
Since this iteration achieved an "Almost perfect agreement" from kappa agreement (except for  "Substantial agreement" for "Sadness" for emotion), there was no need for addition iterations.
Satisfied with the result, all samples were coded, with the first author assigned half of the samples (192), and the rest (192)  evenly divided (64 each).
To evaluate these results, we calculate the accuracy, precision, recall, and F1 score.
Note that since sentiment is not binary, we cannot report on the precision, recall, and f1 score. Instead, we only report accuracy. Furthermore, we also could not report AUROC due to the nature of all tools \footnote{Please refer to \url{https://github.com/NAIST-SE/FirstResponsePR/blob/main/DEVIATIONS.md} for details}.

Table \ref{tab:acc_kappa_sanity} shows the summary of our sanity check, where tools score high accuracy.
This indicates that tool output is consistent with the ground truth. 
The results also indicate that precision and recall are relatively lower. 
A reason for this is due to a skewed distribution in the dataset.
For example, toxicity has a very low precision recall and F1 because the dataset only contains two first responses are classified as "toxic", while 382 were labeled as "non toxic."

\section{Findings}
\label{sec:Finding}

In this section, we present the research motivation, approach and result.

\subsection{Characterizing the First Response (RQ1)}
\label{sec:RQ1}
% \raula{I think it is better to explain the RQ in detail better.\checkmark}

\paragraph{\uline{Motivation.}} Our motivation is to characterize first responses in terms of sentiment, responsiveness, toxicity, and emotion of the response. 
We envision that this analysis will reveal insights on whether there are biases against the first PR.
% \textbf{Hypothesis.} 
We test the following hypothesis:
\begin{itemize}
    \item \textit{H1.1: First contributions are more likely to get positive responses compared to non-first contributions.}  Since the community tries to attract more contributors, we can expect positive responses to first contributions.
    \item \textit{H1.2: First contribution are more likely to get more responsive responses compared to non-first contributions.} Similar to \textit{H1.1}, we expect the community to give responsive responses.
    \item \textit{H1.3: There are biases in giving toxic responses between first contribution and non-first contribution.}  Similar to \textit{H1.1}, we expect the community will be bias in their first response.
    \item \textit{H1.4: Each emotion type is not evenly distributed in first contributions and non-first contributions.} Similar to \textit{H1.1}, we expect emotion in the first response to be different from non-first contributions. For example, there could be a exaggerated 'surprise' emotion when first contributor includes a large-sized contribution when compared to a non-first contribution. 
\end{itemize}

\paragraph{\uline{Approach.}} Our approach is to statistically compare characteristics between first contribution and non-first contribution group (D1). 
To confirm the validity of the statistical analysis, we inspect the data to see whether it is normally distributed or not. Hence, we use we adopt Shapiro-Wilk test \citep{shapirotest} with alpha = 0.05.\footnote{\url{https://docs.scipy.org/doc/scipy/reference/generated/scipy.stats.shapiro.html}} From the test, the distributions of both sentiment score and responsiveness of first responses are significantly different from the normal distribution (p-value $<$ 0.05).
From this distribution, we use non-parametric statistical tests (two-tailed Mann Whitney U test \citep{mann1947} with alpha = 0.05\footnote{\url{https://docs.scipy.org/doc/scipy/reference/generated/scipy.stats.mannwhitneyu.html}}). Additionally, we apply the Cliff’s $\delta$ to evaluate effect size.\footnote{\url{https://ttv1.github.io/cliffsDelta.html}} Effect size is calculated as follows: (1) $|\delta| < 0.147$ as Negligible, (2) $0.147 \leq |\delta| < 0.33$ as Small, (3) $0.33 \leq |\delta| <0.474$ as Medium, or (4) $0.474 \leq |\delta|$ as Large. 
We test the following two hypothesis:
 
 \textit{$H1.1_{null}$: There is no difference between the sentiment score of the first response of first contributions and non-first contributions.}

\textit{$H1.2_{null}$: There is no difference between the responsiveness of the first response of first contributions and non-first contributions.}

The next two hypothesis were tested using Pearson’s chi-squared ($X^2$) \citep{chisqure} as the dependent variables, i.e. toxicity and emotion, are categorical:

\textit{$H1.3_{null}$: Toxicity is evenly distributed in the first response to the first contributions and non-first contributions.}

\textit{$H1.4_{null}$: Each emotion type is evenly distributed in the first responses to the first contributions and non-first contributions.}

\begin{figure}[!]
     \centering
     \begin{subfigure}[b]{0.49\textwidth}
        \centering
        \includegraphics[width=\linewidth]{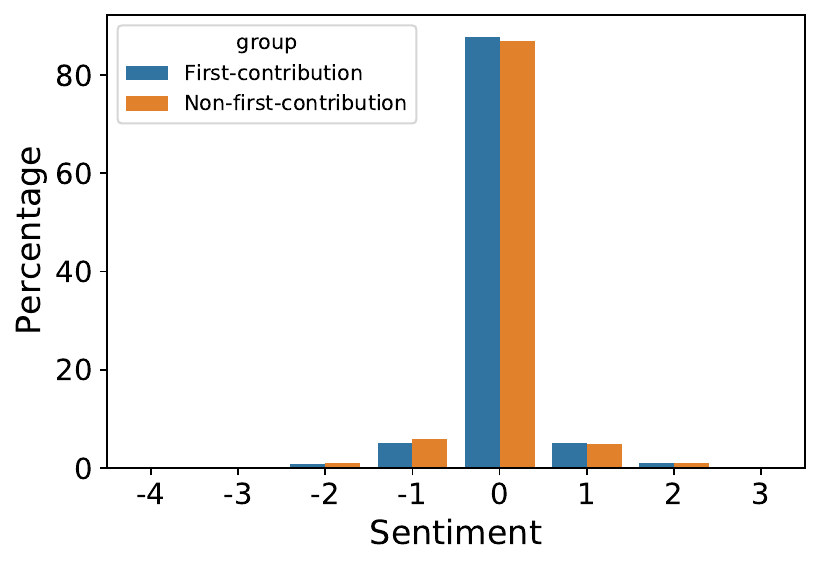}
        \caption{Sentiment Score of the First Response toward First Contributions and Non-First Contributions (\textit{H1.1})}
        \label{fig:rq1_sentiment}
     \end{subfigure}
     \hfill
     \begin{subfigure}[b]{0.49\textwidth}
        \centering
        \includegraphics[width=\linewidth]{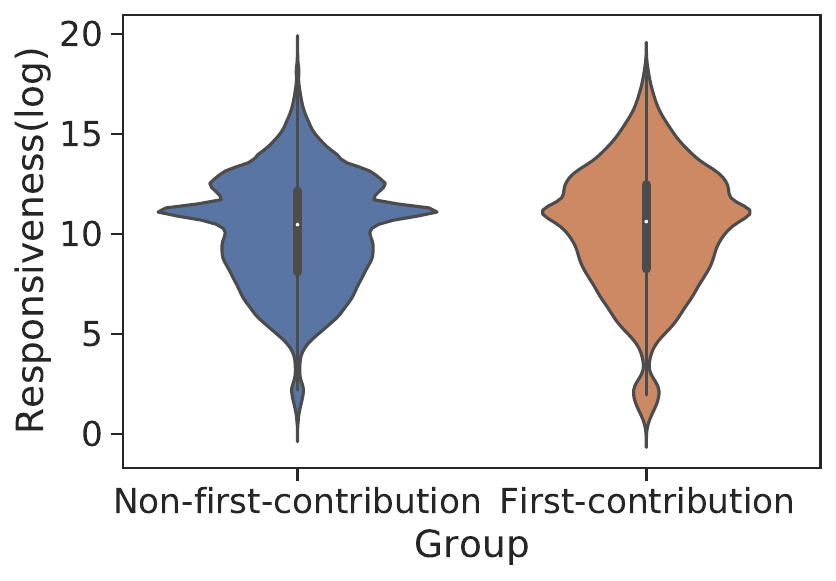}
        \caption{Responsiveness of the First Response toward First Contributions and Non-First Contributions (\textit{H1.2})}
        \label{fig:rq1_responsiveness}
     \end{subfigure}
     \caption{ Visual analysis of Sentiment Score (a) and Responsiveness (b) of the First Response}
\end{figure}
\begin{table}[]
\caption{Toxicity and Emotion statistics in the First Response (\textit{H1.3} and \textit{H1.4})}
\label{tab:rq1_toxic_emotion}
\centering
\begin{tabular}{@{}llrrr@{}}
\toprule
\multicolumn{2}{c}{Tools} & \multicolumn{1}{c}{First Contribution} & \multicolumn{1}{c}{Non-First Contribution} & \multicolumn{1}{c}{Odds Ratio} \\ \midrule
Toxicity  \% & \multicolumn{1}{l|}{} & 0.25 & 0.26 & 0.95\\
\multirow{6}{*}{Emotion} & \multicolumn{1}{l|}{Anger \%} & 1.34 & \textbf{1.50} & 0.90\\
 & \multicolumn{1}{l|}{Fear \%} & \textbf{8.04} & 7.82 & 1.03\\
 & \multicolumn{1}{l|}{Joy \%} & \textbf{7.15} & 6.44 & 1.12\\
 & \multicolumn{1}{l|}{Love \%} & \textbf{1.97} & 1.79 & 1.10\\
 & \multicolumn{1}{l|}{Sadness \%} & 2.77 & \textbf{2.93} & 0.95\\
 & \multicolumn{1}{l|}{Surprise \%} & 13.34 & \textbf{16.70} & 0.77\\ \bottomrule
\end{tabular}
\\
\scriptsize
*The bold indicated the higher detected percentage of toxicity/emotion in the first responses.
*Odds Ratio is calculated by considering odd that First Contribution will receive <<emotion>> response compare to Non-First Contribution.
\end{table}

\paragraph{\uline{Findings.}}
Figure \ref{fig:rq1_sentiment} and Figure \ref{fig:rq1_responsiveness} shows the sentiment score and responsiveness of first response for both first and non-first contributions ($H1.1$ and $H1.2$).
To view our results easily, we present the responsiveness on a log scale, with the wait time measured in days.
Findings indicate first contributions received a higher sentiment score of first response compared to non-first contributions. However, first contributions waited longer to receive first response.
From Figure 3a, we observe that non-first contributions receive slightly lower sentiment score in the distribution compared to first contributions, and responsiveness are 10.953 and 6.463 days on mean average for the first and non-first contributions consecutively as shown in Figure 3b.

Complementary,  Table \ref{tab:rq1_toxic_emotion} shows the number of first response classified as Toxic and Non Toxic, and <<emotion>> and Non <<emotion>> by the aforementioned tools ($H1.3$ and $H1.4$). 
We see that first contributions contain lower percentage of "Toxic" response when compared to non-first contributions. On one hand for emotion, percentages of "Fear", "Joy", and "Love" first responses are higher in first contributions, while percentages of "Anger","Sadness", and "Surprise" first responses are lower.

Statistically, we confirm the significant differences in the sentiment score and responsiveness of first responses of first and non-first contributions, rejecting the null hypothesis since p-value $<$ 0.05 for both $H1.1_{null}$ and $H1.2_{null}$. 
Also, this effect was negligible  $|\delta|$ of 0.009 (negligible) and 0.043 in both cases. 
Additionally, we reject the null hypothesis $H1.4_{null}$ (p-value < 0.05)  that all of the emotion types are not evenly distributed in first responses, and $H1.3_{null}$ not being significant. 

\begin{tcolorbox}
    \textbf{Answering RQ1:}
    Results indicate that first contributions are more likely to be positive (\textit{H1.1}).
    First contributions are less responsive (\textit{H1.2}) when compared to non-first contributions. 
    Although toxic responses are not bias to the first contribution (\textit{H1.3}), but developers did express specify sentiments (i.e., fear, joy and love) in the first response (\textit{H1.4}).
\end{tcolorbox}

\subsection{First response affecting future interactions (RQ2)}
\label{sec:RQ2} 
\paragraph{\uline{Motivation.}} Our motivation is to understand whether first responses affect future interactions between first-time contributors and project contributors.
We expect to reveal the effects of the first responses and test these two hypothesis:
\begin{itemize}
    \item \textit{H2.1:} First-time contributors are more likely to interact with non-first-time contributors after the first responses when they receive positive responses.
    Receiving positive responses encourage the first-time contributors' willingness to collaborate \citep{Grigore_2011} in the future.
    \item \textit{H2.2:} First-time contributors are more likely to interact with non-first-time contributors after the first response.
    A response can keep the first-time contributors' attention, i.e., the first-time contributors are not waiting, causing first-time contributors to be interactive with others.
\end{itemize}
Furthermore, we manually inspect and characterize negative first responses.

\paragraph{\uline{Approach.}} Our approach includes both a quantitative and qualitative components.
Using dataset D2,  we compare the sentiment and responsiveness of first responses between contributions that first-time contributors interacted with project contributors against contributions that first-time contributors did not interact with project contributors.

\textbf{Quantitative Analysis.}
Similar to \textit{H1.1} and \textit{H1.2}, we use the same protocol to test the following:

\textit{$H2.1_{null}$: There is no difference between the sentiment score of the first response of the first-time contributors that interacted with project contributors and one that did not interact}

\textit{$H2.2_{null}$: There is no difference between the responsiveness of the first response of the first-time contributors that interacted with project contributors and one that did not interact.}

\textbf{Qualitative Analysis.} We conduct a qualitative analysis to understand what is contained in the negative response.
To understand the intention behind these responses, we split the classification into two categories, and perform a systematic coding, similar to prior work \cite{Wang2021}.
 After conducting two iterations of coding 30 samples, we classify each response based intention and the emotion communicated between during the interaction. 
 Coding was performed together, with agreements negotiated.
 After two iterations, we decided upon these two taxonomies, as shown below:
 
 \texttt{Taxonomy One: Literal Intention of Message} which does not include emotion. Coding with examples includes:
\begin{itemize}
    \item \textbf{Suggestion to author:} In this case, the response was intended to provide the author with concrete advice in terms of problem solving or additional work.
    \begin{quote}
    \textit{ex. looking at some of the examples at <<url>>, i'm inclinded to think that, in this case, <<script>> should be implemented as <<script>> within the anchor (as much as i hate to add more spans)}
    \end{quote}
    \item \textbf{Question to author:} In this case, the reviewer asked a question with the intention for the author to reply back.
    \begin{quote}
        \textit{ex. weird. it works but should not it be <<script>> ?}
    \end{quote}
    \item \textbf{Comments Only: } A comment  that has no question or suggestion.
    \begin{quote}
        \textit{ex. allowing the attacker to edit/create files <<filename>> by root. this exploit}
    \end{quote}
    \item \textbf{Other:} cases that do not fit into the other categories.
    \begin{quote}
        \textit{ex. oops}
    \end{quote}
\end{itemize}
\texttt{Taxonomy Two: Emotion arousal of Message} which includes the emotion that the response intends to communicates. Coding with examples include: 
\begin{itemize}
    \item \textbf{Ironic:} The response uses language that normally signifies the opposite. This is typically for humorous or emphatic effect.
    \begin{quote}
        \textit{ex. this is absolutely nitpicking but i find the presence of a quotation mark in a doc string so *weird*!}
    \end{quote}
    \item \textbf{Constructive:} Although at first glance it does seem negative and harsh, in the context of the discussion, it may invoke negative emotions but the intention is to be helpful.
    \begin{quote}
        \textit{ex. could you check for the error case before calling <<script>> so you do f have to catch deverror? also would be good to have a test}
    \end{quote}
    \item \textbf{Apologetic:} The response is intended to communicate an apologetic nature, and invoke empathy.
    \begin{quote}
        \textit{ex. sorry about that, should not have dropped you out of the label. i will merge in a change putting you back as the author.}
    \end{quote}
    \item \textbf{Criticizing:} Different to constructive, the response is not meant to help the author, and can invoke negative emotions and does not imply to be helpful.
    \begin{quote}
        \textit{ex. mistake?}
    \end{quote}
    \item \textbf{Figurative:} The response contains emojis to communicate feelings that are beyond descriptions and does not fit into any other categories.
    \begin{quote}
        \textit{ex. this is not going to work :( <<script>> (and <<script>>) do not depend on mock, and although it is being added to <<filename>> as a dependency it will break package workflows that require declaring dependencies for running tests. instead, we can use <<script>>}
    \end{quote}
    \item \textbf{Other:} False positives or cases that do not fit into the other categories.
    \begin{quote}
        \textit{ex. you removed apache announcement and retained cheat sheet}
    \end{quote}
\end{itemize}

Similar to other software engineering studies \cite{asia_2020} and the sanity check, we adopt use the Kappa's agreement.
Three authors separately coded the first 30 samples of negative first response.  With a “almost perfect" score of 0.84 (literal intention) and 0.81 (emotion arousal), then proceeded to equally distribute and code the rest of the samples. 

\begin{figure}[!]
     \centering
     \begin{subfigure}[b]{0.49\textwidth}
        \centering
        \includegraphics[width=\linewidth]{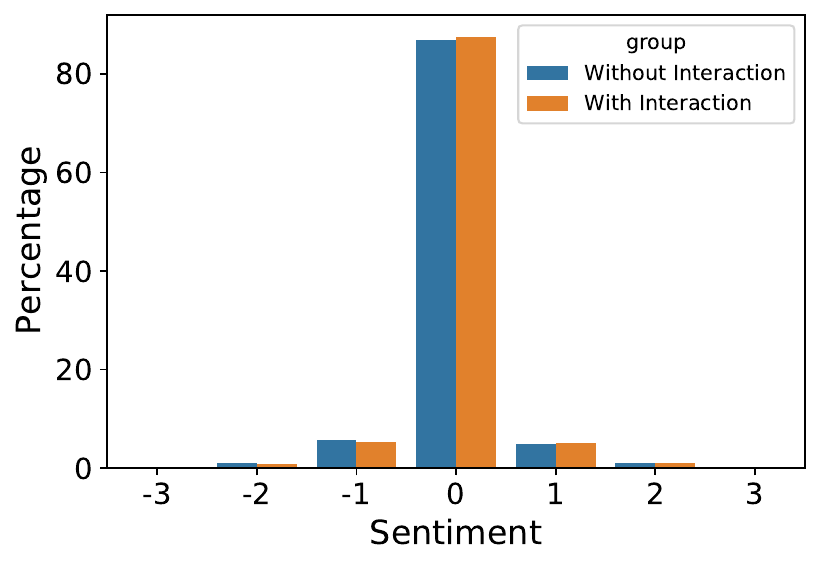}
        \caption{Sentiment Score of the First Response toward First Contributions\textit{H2.1}}
        \label{fig:rq2_sentiment}
     \end{subfigure}
     \hfill
     \begin{subfigure}[b]{0.49\textwidth}
        \centering
        \includegraphics[width=\linewidth]{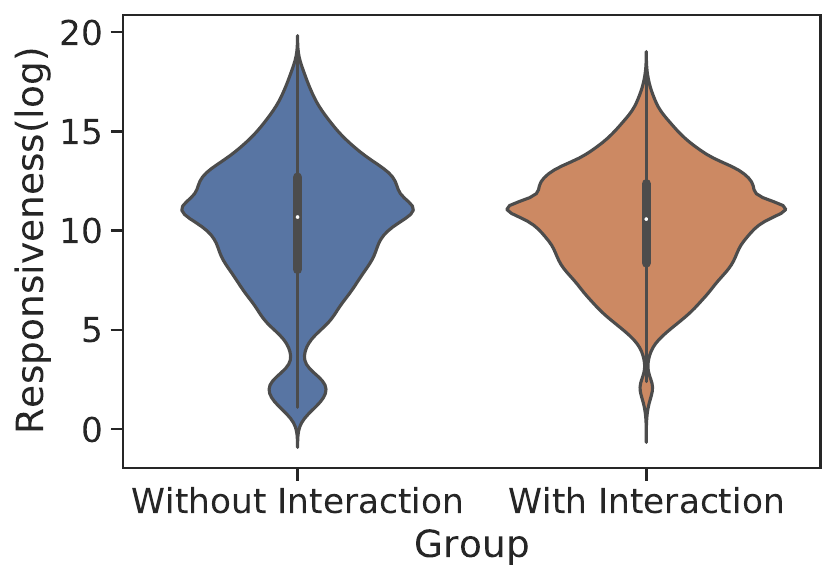}
        \caption{Responsiveness of the First Response toward First Contributions \textit{H2.2}}
        \label{fig:rq2_responsiveness}
     \end{subfigure}
     \caption{Quantitative results to answer RQ2}
\end{figure}

\begin{table}[]
\caption{Distribution of the Qualitative Analysis to answer RQ2}
\label{tab:rq2_qualitative}
\centering
\begin{tabular}{@{}lll@{}}
\toprule
\multicolumn{1}{c}{Taxonomy} & \multicolumn{1}{c}{Category} & \multicolumn{1}{c}{Percentage} \\ \midrule
\multirow{4}{*}{Literal Intention} & Suggestion to Author & 44.36 \mybar{0.44}\\
 & Question to Author & 30.71 \mybar{0.31}\\
 & Comment Only & 32.28 \mybar{0.32}\\
 & Other & \phantom{0}1.84 \mybar{0.02}\\ \midrule
\multirow{6}{*}{Arousal Emotion Intended} & Ironic & \phantom{0}2.62 \mybar{0.03} \\
 & Constructive & 51.97 \mybar{0.52}\\
 & Apologetic & \phantom{0}3.41 \mybar{0.03}\\
 & Criticizing & 37.53 \mybar{0.38}\\
 & Figurative & \phantom{0}1.84 \mybar{0.02}\\
 & Other & \phantom{0}7.09 \mybar{0.07}\\ \bottomrule
\end{tabular}
\end{table}

\paragraph{\uline{Findings.}} Figure \ref{fig:rq2_sentiment} and Figure \ref{fig:rq2_responsiveness} shows the sentiment score and responsiveness of first responses toward first contributions that the first-time contributors interacted and did not interact. We observe that mean responsiveness are 7.080 
and 6.953 days. For both $H2.1_{null}$ and $H2.2_{null}$, they were statistically not significant.

For our qualitative analysis, Table \ref{tab:rq2_qualitative} show the result from our manual coding. We find that there are biggest message conveyed were suggestions (44.36\%) and comments (32.28\%). For the emotion aroused, the two most conveyed was constructive (51.97\%) and criticizing (37.53\%).
Others were less prevalent (7.09\% - other, 3.41\% - apologetic, 2.62\% - ironic, and 1.84\% -figurative).
\begin{tcolorbox}
    \textbf{Answering RQ2:}
    Results did not find any statistical relationships between a positive response (\textit{H2.1}) and a responsive response to increase the likelihood for a future interaction (\textit{H2.2}).
    Further investigation reveal that negative first responses include suggestions (51.97\%), comments (32.28\%) and questions (30.71\%), and have constructive (51.97\%) or criticizing (37.53\%) attributes.
    % \raula{add the \%s}
%     Our statistical testing cannot confirm that first-time contributors are more likely to interact when they receive positive responses (\textit{H2.1}). In the same way, our result from statistical testing cannot confirm the relationship between interactions of first-time contributors and receiving responsive responses (\textit{H2.2}).
%     We identified 6.01\% of first responses towards first contributions are negative. Most of the intentions in the negative first response are comment, suggestion, and question. The majority of the rationale for the negative first response is to provide constructive feedback to the first-time contributors. 
\end{tcolorbox}

\subsection{Factors affecting future contributions (RQ3)}
\label{sec:RQ3}

\paragraph{\uline{Motivation.}} Our motivation is to reveal insights into factors that affect the future contributions.

\paragraph{\uline{Approach.}} Our approach is quantitative and uses dataset D4. 
% Answering this research question will reveal a correlation between interactions between first-time contributors and project contributors, and future contributions. 
Since we frame this problem as binary classification problem, we train classification models followings \cite{Tuarob_2021}. The models train on independent variables by leveraging dependent variables that are introduced below:
\begin{itemize}
     \item \textbf{Sentiment} (\texttt{sentiment}): Our rationale is that certain sentiments may correlate with a future contribution. For example, a positive interaction with a first time contributor might encourage a future contribution. 
    % We extract this variable by using SentiStrength-SE \citep{sentiment_tool}. 
    \item \textbf{Responsiveness} (\texttt{responsiveness}): Our rationale is that a more responsive interaction may correlate with a future contribution. For example, newcomers might be willing to make a contribution based on a fast responsiveness on their submitted PRs. 
    % This variable is calculated as the duration between the PR and the first response creation date.
    \item \textbf{\# Interactions} (\texttt{interaction}): Our rationale is that the number of interactions in the current PR may correlate with the future contribution. For example, higher number of interaction potentially shows that the contributor is active and could lead to future contribution.
    % We calculate this variable by considering the total number of comments on a PR. More specifically, we count the number of all comments that were written after the first response of the first-time contributors on the PR.
    \item \textbf{\# Words in Response} (\texttt{word\_first\_response}): Our rationale is that the the length of the first response may correlated with future contribution. For example, Longer first response could contain details to help understand feedback, resulting in future contributions. 
    % This variable is the total words in the first response. 
    \item \textbf{Outcome of Review} (\texttt{merged}): Our rationale is that the result of the current PR may affect the future contribution. For example, the result of PR could increase the possibility of future contribution. This is a binary variable that shows whether the PR was merged or not. 
    \item \textbf{Toxicity in Response} (\texttt{toxic}): Our rationale is that the existence of toxicity in the response may negatively affect the future contribution. For example, the response that contains toxic could be a barrier to new contribution. 
    % This is the binary variable that shows whether the first response contains toxicity or not. 
    \item \textbf{Emotion in Response} (\texttt{<<emotion>>}): Our rationale is that certain emotion in response could impact the possibility of future contribution. For example, the contributor could leave the project after receiving reviewing response that contains anger emotion. 
    % As the EmoTxt allows us to detect each emotion separately, we will consider six binary variables which describe the emotion in the first response. For example, "Anger" will be a binary variable in the first response. The emotions consist of "Joy", "Anger", "Sadness", "Love", "Surprise", and "Fear." 
\end{itemize}
%  
% 
% Since contributing to OSS projects contains much more dimensions of factors than the interaction dimension, we incorporate other dimensions of factors as well.
Contributors Dimension variables captures the following:
\begin{itemize}
    \item \textbf{\# PRs} (\texttt{other\_project\_pr}): Our rationale is that PR experience increases continued contribution. 
    \item \textbf{Programming language} (\texttt{other\_same\_language\_pr}): Our rationale is that specific programming language might have an impact.
    As same as \texttt{other\_project\_pr}, this variable is calculated according to the time that the current PR was created.
    \item \textbf{GitHub Experience} (\texttt{gh\_account\_duration}): Our rationale is that more experienced users of GitHub are likely to make more contributions. Age is calculated according to when the account was first created.
\end{itemize}
Project Dimension variables captures the following:
\begin{itemize}
    \item \textbf{\# commits }(\texttt{project\_commit}): Our rationale is larger projects draw more contributions.
    \item \textbf{\# PRs} (\texttt{project\_pr}): Similar to  \texttt{project\_commit} we measure by number of PRs instead of number of commits. 
    \item \textbf{Programming Language} (\texttt{language\_<<Programming language>>}): Our rationale is a contributors difference to the project programming langauge may impact future contributions.  
\end{itemize}
Contribution Dimension variables captures the following:
\begin{itemize}
    \item \textbf{PR's commit} (\texttt{pr\_commit}): Our rationale is that the size of contribution impacts future contribution. We count the commits per PR. 
\end{itemize}

Additionally we calcuate the \texttt{Future Contribution} dependent variable, as a binary variable to identify first-time contributors that do make a future contribution. To filter out inactive contributors (first-time contributors that have left the project) , set  maximum days as a threshold.

\paragraph{Selecting a Machine Learning model.} 
Table \ref{tab:mean_performance_r3} shows a summary of the machine learning models used. 
Following \cite{Tuarob_2021}, we compare well-known machine learning models of Logistic Regression (LR), Decision Tree (DT), Random Forest (RF), SVM, Naive Bayes (NB), QDA, K Nearest Neighbors (KNN), and Neural Network (NN). Additionally, we added dummy model (random) to the comparison as a baseline for the model. We apply default parameters \footnote{\url{https://scikit-learn.org/stable/}} and a 10-fold cross-validation.
Random Forest performing the best in terms of the precision, recall, F1 score, and AUROC (Area Under the ROC Curve). Note that the precision, recall, and F1 score are in regard to positive classes. It is also worth noting that SVM and NB have higher recall and precision than RF respectively.
To validate this result, we use the two-tailed Mann Whitney U test and Cliff’s $\delta$ effect size.
We confirm RF is most suitable with the highest difference and effect size as shown in Table \ref{tab:statistical_rq3}.

\begin{table}[]
\caption{Average Performance of Each Machine Learning Model}
\label{tab:mean_performance_r3}
\centering
\begin{tabular}{@{}lrrrr@{}}
\toprule
\multicolumn{1}{c}{Model} & \multicolumn{1}{c}{Precision} & \multicolumn{1}{c}{Recall} & \multicolumn{1}{c}{F1} & \multicolumn{1}{c}{AUROC} \\ \midrule
Dummy                        & 0.5041                        & 0.5028                     & 0.5035                 & 0.5046                    \\
LR                        & 0.5863                        & 0.6071                     & 0.5965                 & 0.6152                    \\
DT                        & 0.5583                        & 0.5436                     & 0.5508                 & 0.5750                    \\
RF                        & 0.6212                        & 0.6130                     & \textbf{0.6171}        & \textbf{0.6671}           \\
SVM                       & 0.5719                        & \textbf{0.6667}            & 0.6157                 & 0.6157                    \\
NB                        & \textbf{0.6331}               & 0.2399                     & 0.3480                 & 0.6004                    \\
QDA                       & 0.6033                        & 0.3069                     & 0.4068                 & 0.5872                    \\
KNN                       & 0.5567                        & 0.5428                     & 0.5496                 & 0.5703                    \\
NN                        & 0.5621                        & 0.6475                     & 0.6018                 & 0.5949                    \\ \bottomrule
\end{tabular}
\\
\scriptsize The bold indicated the model that perform best according to each metrics
\end{table}
\begin{table}[]
\caption{A comparative summary of the predictive accuracy between RF and other machine learning models. The bold
text indicates that RF is better than other models.}
\label{tab:statistical_rq3}
\begin{tabular}{@{}rrrrrrrrr@{}}
\toprule
\multirow{2}{*}{\begin{tabular}[c]{@{}r@{}}RF vs \\ Baselines\end{tabular}} & \multicolumn{2}{c}{Precision}                             & \multicolumn{2}{c}{Recall}                                & \multicolumn{2}{c}{F1}                                    & \multicolumn{2}{c}{AUROC}                                 \\
                                                                            & \multicolumn{1}{c}{\%Diff} & \multicolumn{1}{c}{Eff.Size} & \multicolumn{1}{c}{\%Diff} & \multicolumn{1}{c}{Eff.Size} & \multicolumn{1}{c}{\%Diff} & \multicolumn{1}{c}{Eff.Size} & \multicolumn{1}{c}{\%Diff} & \multicolumn{1}{c}{Eff.Size} \\ \midrule
Dummy                                                                          & \textbf{18.85}             & L***                         & \textbf{17.98}             & L***                         & \textbf{18.41}             & L***                         & \textbf{24.36}             & L***                         \\
LR                                                                          & \textbf{5.62}              & L***                         & 0.96                       & $\circ$                            & \textbf{3.34}              & L***                         & \textbf{7.78}              & L***                         \\
DT                                                                          & \textbf{10.13}             & L***                         & \textbf{11.32}             & L***                         & \textbf{10.74}             & L***                         & \textbf{13.81}             & L***                         \\
SVM                                                                         & \textbf{7.94}              & L***                         & -8.76                      & L***                         & 0.23                       & $\circ$                            & \textbf{7.70}              & L***                         \\
NB                                                                          & -1.92                      & L*                           & \textbf{60.86}             & L***                         & \textbf{43.61}             & L***                         & \textbf{10.00}             & L***                         \\
QDA                                                                         & 2.88                       & $\circ$                            & \textbf{49.93}             & L***                         & \textbf{34.08}             & L***                         & \textbf{11.98}             & L***                         \\
KNN                                                                         & \textbf{10.38}             & L***                         & \textbf{11.45}             & L***                         & \textbf{10.94}             & L***                         & \textbf{14.51}             & L***                         \\
NN                                                                          & \textbf{9.51}              & L***                         & -5.63                      & $\circ$                            & 2.48                       & $\circ$                            & \textbf{10.82}             & L***                         \\ \bottomrule
\multicolumn{9}{l}{\multirow{2}{0.95\linewidth}{\scriptsize \textbf{Effect Size:} Large (L) $r > 0.5$, Medium (M) $0.3 < r \leq 0.5$, Small (S) $0.1 < r \leq 0.3$, Negligible (N) $ r < 0.1$}} \\ \\
\multicolumn{9}{l}{\scriptsize \textbf{Statistical Significance:} $^{***}p < 0.001, ^{**}p < 0.01, ^*p < 0.05, \circ p \geq 0.05$}
\end{tabular}

\end{table}

\paragraph{Feature Importance.} 
To identify correlated variables with a future contribution, we perform two analysis. 
The first is the feature importance analysis\footnote{\url{https://scikit-learn.org/stable/modules/permutation_importance.html}} was using the RF model. 
Using a 10 iteration feature shuffling, we compare the feature importance between the first-time contributors' interaction variables against the rest of the variables.
The second analysis is a partial dependence plot between top important variables and \texttt{Future Contribution}.\footnote{\url{https://scikit-learn.org/stable/modules/partial_dependence.html}}.
This will show how each variable is associated with the future contribution.

\begin{figure*}[!t]
\centering
\includegraphics[width=.8\linewidth]{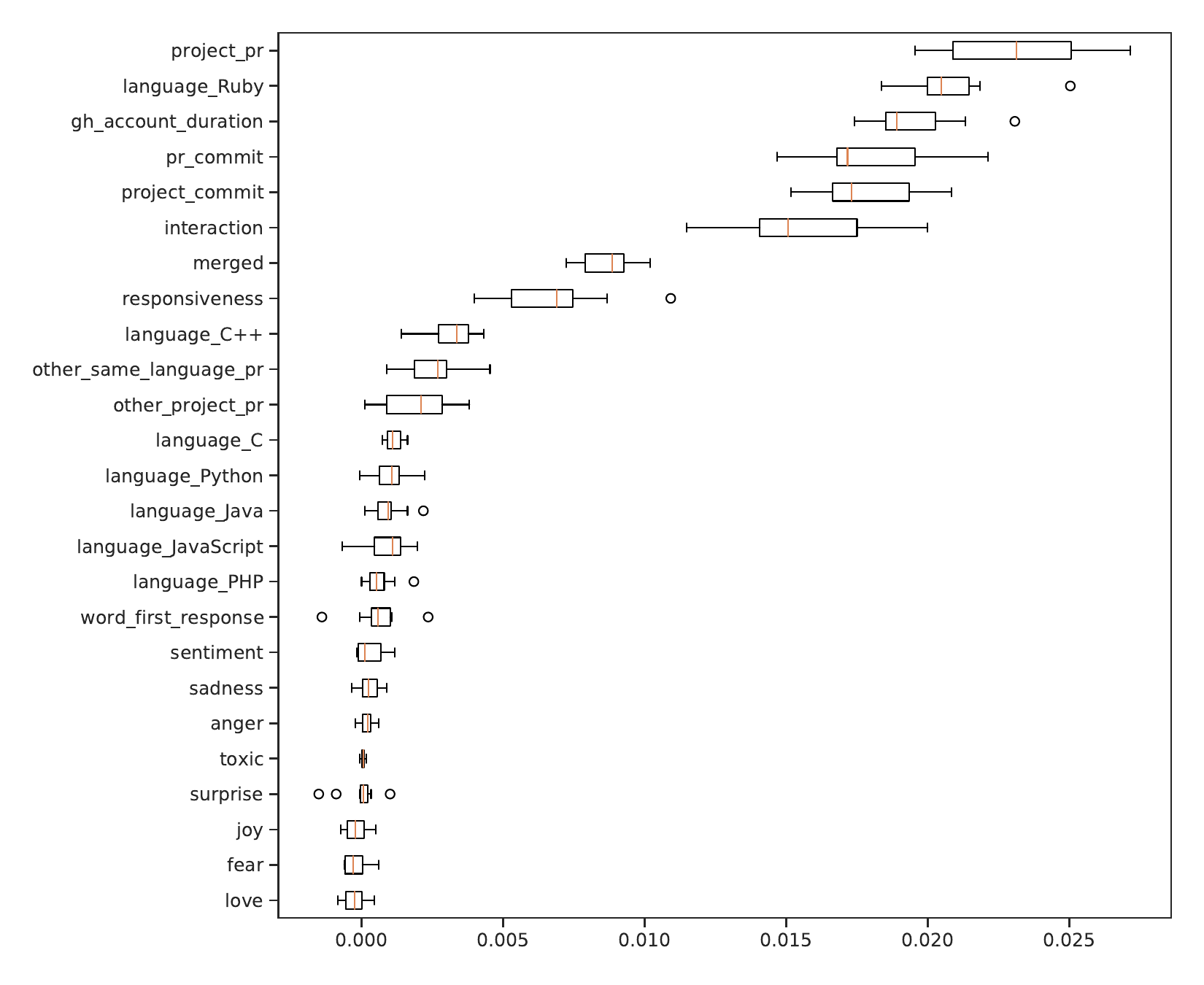}
\caption{Feature Permutation Importance for RQ3}
\label{fig:rq3_FPT}
\end{figure*}

\textbf{Findings.} Figure \ref{fig:rq3_FPT} shows the feature permutation importance by using 10 iteration of feature shuffling. 
Our RF model indicates that  \texttt{project\_pr}, \texttt{language}, \texttt{gh\_account\_duration},\texttt{pr\_commit}, \texttt{project\_commit}, and \texttt{interaction} as the most important features.
Interestingly, we find that compared to the \texttt{interaction} variable, there are three project dimension, one contributor dimension, and one contribution dimension variables that are more important. 

\begin{figure*}[!t]
\centering
\includegraphics[width=\linewidth]{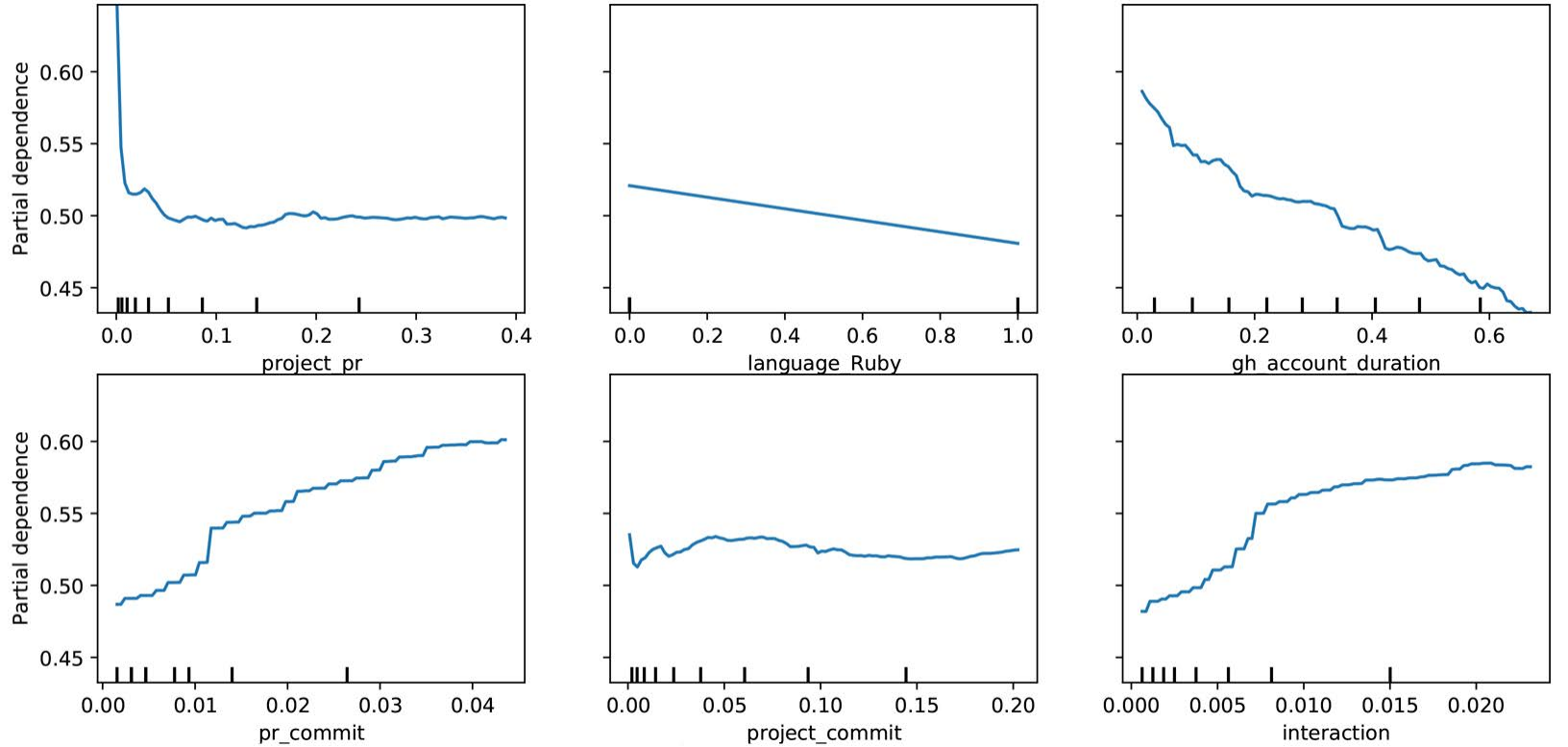}
\caption{Variables Partial Dependence Plots}
\label{fig:rq3_pdp}
\end{figure*}

% Figure \ref{fig:rq3_pairwisecorr} shows the pair-wise correlation between each variable and \texttt{Future Contribution}. The result shows that the \texttt{interaction} show strongest positive correlation to \texttt{future contribution}. \texttt{Responsive}, which is duration needed to wait for the first response, is negatively correlated to \texttt{future contribution} while \texttt{sentiment} is positively correlated to \texttt{future contribution}. However, the correlation is not very high compared one of the \texttt{interaction}. This implies that interactions between first-time contributors and project contributors are more important than having positive and a quick response.
% % This imply that receiving positive and fast first response is associated with likelihood of future contribution.

% \begin{figure*}[!t]
% \centering
% \includegraphics[width=\linewidth]{pairwise_correlation.pdf}
% \caption{Variable Pair-wise Correlation}
% \label{fig:rq3_pairwisecorr}
% \end{figure*}

Figure \ref{fig:rq3_pdp} shows the partial dependence plot between six of the most important variables (\texttt{project\_pr}, \texttt{language}, \texttt{gh\_account\_duration},\texttt{pr\_commit}, \texttt{project\_commit}, and \texttt{interaction}). 
Out of these six variables, we see the \texttt{interaction} and \texttt{pr\_commit} have a positive correlation to \texttt{future contribution}.

\begin{tcolorbox}
    \textbf{Answering RQ3:} From our prediction models, results indicate that contributor interactions have positive correlation with having a future contribution.
    In terms of the model, we identify several important predictive features related to project (\# PR, \# commit , language),  contributor (GitHub account), and  contribution (size).
\end{tcolorbox}

\section{Protocol Implications vs. Study Findings}
\label{sec:implication}
Referring back our expected registered report implications \citep{Assavakamhaenghan:msr2021}, we now compare against our actual findings.

\textbf{Suggestions for OSS Projects.}
Our expected implications was to provide insights into how OSS projects affects future contributions. We assumed that fast and appropriate responses to first-contributors are needed to attract future contributions. 
In contrast, our findings contrast our expectations as well providing new insights.
First, RQ1 findings suggest that OSS projects are already positive, and express fear, joy or love in the responses.
% This means that OSS projects are already are positive. 
Second, from the findings of RQ2, we report that positive and fast responses is not significantly correlated to further interaction in the first contribution.
Furthermore, most of the negative sentiments expressed in the first-response is either constructive or criticizing. 
We speculate that to conform with standards of the projects, they may need to provide negative but constructive comments, which might not have the intention of being toxic.
With these findings, our recommendation for OSS projects is to continue with their constructive comments if needed. 
As raised by prior work, as long at the negative sentiment is not perceived as being toxic, they should not have any issues.

\textbf{Suggestions for Researchers.}
Our expected implication was that the study would reveal interaction-related factors matter for future contributors and to what extent does the factors affect OSS projects' contributions. 
From RQ2, we find that negative responses are constructive and criticizing comments and suggestions.
Since the findings from RQ3 findings shows do not have a effect on future contributions, with low performance, it might mean that either existing detection tools are not yet adequate, or that the variables that we select are not exhaustive. 
We show that the toxicity tools may not be as efficient (shown by the sanity check), thus a potential future avenue of research is improved tool support, and taking into account new variables with improved models. 
This should be future work.

\textbf{Suggestions for Contributors.}
We expected from the registered protocals that contributors would understand the importance of their interactions. 
Overall, all findings indicate that the sentiment should not be seen as a social barrier for new contributors. 
Thus, we advise that contributors should embrace the negative comments and suggestives as being constructive, that will be useful feedback for the contributor. 
Contributors should instead strive to read between the sentiment to take the constructive messages to improve their chances of their contributions from getting accepted. 
Furthermore, contributors should not be afraid of first-responses as most are positive, and show sentiments of fear, joy and love.
    
\section{Threats to Validity}
\label{sec:threats}
In this section, we discuss the threats to the validity of our study including external, construct, and internal validity.

\textbf{External Validity} refer to the generalization of our results, and the external tools. A key threat we acknowledge is the accuracy of the detection tools as no tool is perfect. To mitigate this risk, we perform a sanity check.
Secondly, our results may only apply to GitHub projects, however, being one of the most popular OSS hosted platform, our results does cover many first-contributors.
Also, we do not consider other forms of communication such as email, discussions, or an issue system. 

\textbf{Construct Validity} refer to the concerns between the theory and our results. A key threat to validity is low model performance.
Predicting whether or not a contributor will make a future contribution based on the emotions, is not an easy Software Engineering phenomena, as there may be several factorsnot included in our scope. By including a baseline, we do claim that our models still perform relatively better than a random coin-toss. To mitigate the threat of multiple identity, we employ quality checks, including removing bots (bots). 
To mitigate threats relating to the quality dataset, we (1) removed PRs with no response and made sure to (2) leave the longest time window so that we could capture any potential future contributions. We also acknowldge that a response written in non-English language could affect our results. Lastly, incorrect prediction and classifications could lead to the incorrectness in the result of our hypothesis testing, but we believe the sanity check does give us confidence to mitigate this threat.

\textbf{Internal Validity} refer to the concerns with the approach taken to execute the study. We acknowledge the feature permutation importance is based on only a single machine learning model (random forest), and may not be consistent for the other models. Furthermore, we acknowledge limitations. However, we our approach with multiple machine learning models provides the best performing systematically.

\section{Related Work}
\label{sec:Related_Work}
% In this section, we discussed our related work.
% \raula{we need to clean up the related work}

\textbf{Sustainability of Open Source Software.} 
We divide sustainable OSS into three areas.
The first topic is ensuring sustainability in software.
Ensuring sustainability is a persistent challenge in
OSS communities \citep{GAMALIELSSON:jss2014}.
% Understanding the motivations behind why developers make contributions to OSS projects is vital to their sustainability.
\cite{Hars:2001} revealed that the common motivations to make contributions to OSS projects are the joy of programming, the identification with a community, career advancement and learning.
\cite{Roberts:2006} explored the interrelationships between OSS developer motivations, revealing the motivations are not always complementary.
When comparing motivations between individual developers and companies, \citet{Bonaccorsi:2006} observed that companies are more motivated by economic and technological reasons.
\citet{Lee:icse2017} found that the most common motivation of one time contributors is to fix bugs affecting their works, while the highly mentioned motivation of casual contributors is ``scratch their own itch'' \citep{Pinto:saner2016}. Compared to our work, we further investigate the motivation of first-time contributors in continuing their contribution in four difference dimensions of factors.

The second area studies the barriers to sustainable software.
% Prior works also revealed the barriers and challenges faced by developers when contributing to OSS projects to assist the developer onboarding and participation.
A study by \citet{Fagerholm:esem2014} observed that mentoring increases the chance of developers’ active participation.
In the pull-based model, a survey by \citet{Gousios:icse2016} revealed that the most commonly reported challenge is the lack of responsiveness from project integrators.
%\citet{Steinmacher:chase2013} investigated first interactions of newcomers on a OSS project. Their results show that kinds and authors of received answer impacted on the onboarding of newcomers.
\cite{Steinmacher:icse2018} analyzed PRs of quasi-contributors and found that the nonacceptance demotivated or prevented them from placing another contributions.
\citet{Li:tse2021} observed the significant impacts of PR abandonment on project sustainability. In the same way, our work studied whether or not the first response toward first contributions is a barrier for first-time contributors.

The third area of related work is the willingness of developers to sustain contributions.
% Additionally, to remain long-term and established OSS contributors, prior works explored the factors affecting the willingness of developers.
A study by \citet{Zhou:icse2012} found that developers’ willingness and participation environment significantly impact on the chance of becoming long-term contributors.
\citet{Schilling:hawaii2012} showed that developers’ retention in OSS projects affected by the level of development experience and conversational knowledge.
%A survey of \cite{Miller:oss2019} with disengaged OSS contributors showed that occupational problems, e.g., changing to a new job that does not support OSS development work, are the most common reasons.
\citet{Damien:2018} studied the impact of PR decisions on future contributions and found that continued contribution to a project is correlated with higher PR acceptance rates.
\citet{Iaffaldano:IEEE2019} conducted an interview with OSS developers to explore what drives them to become temporarily or permanently inactive in a project.
% The reported reasons were personal (e.g., life events) or project related (e.g., role change and changes in the project).
Compared to these studies, we found that interaction are the most important factor for future contribution.

\textbf{Newcomers and their first contributions.} 
% We divide related work into three different areas.
% The first area is the motivation for newcomer to contribute to OSS.
Newcomers play an important role for OSS survival,
long-term success, and continuity \citep{Kula:2019}.
% In order to attract and retain newcomers, previous studies investigated the motivations behind their joining in OSS projects.
\citet{Lakhani:2003} revealed that new contributors are primarily motivated by external benefits (e.g., better work, career advancement), followed by fun, code-based challenges, and improved programming skills.
\citet{Choi2010} showed that the natural decline of newcomer editing could be delayed by a welcome message, technical assistance and constructive criticism.
% Meanwhile, these studies explored how newcomers join projects in order to become core contributors \citep{Fang:2009, Marlow:2013}.
A recent study by \citet{Subramanian2022} characterized contributions attract newcomers. This contrast to our work as we find that the positivity and responsiveness is not necessary for first-time contributor interaction.

% The second area of research is related to the mentorship to support a newcomer.
Prior works in this area also pay attention to the mentorship to support the newcomer onboarding.
\citet{Swap:2001} defined the mentorship as a basic knowledge transfer mechanism in the enterprise.
\citet{Sim:1999} found mentoring patterns when new developers join software projects. 
Through an empirical study of OSS projects, \citet{Nakakoji:2003} proposed eight possible joining roles, consisting of concentric layers called "the onion patch".
\citet{VONKROGH:2003} proposed a joining script for developers who is willing to participate in a project.
Compare to these study, our study instead investigate the first responses toward first-time contributor instead of mentorship between developer. 
% We also analyze the correlation between the interaction and future contribution.

% The third area of research are barriers that newcomers face. 
% \citet{Steinmacher:chase2013} investigated first interactions of newcomers on a OSS project. Their results showed that kinds and authors of received answer impacted on the onboarding of newcomers.
% \citet{Steinmacher2014} proposed a developer joining model, representing common stages and forces that impact on newcomers being drawn or pushed away from a project.
% To support newcomers, they also proposed a portal called FLOSScoach based on a conceptual model of barriers \citep{Steinmacher:icse2016}. Results showed that the portal played an important role in guiding newcomers and lowering orientation and contribution process-related barriers.
% Furthermore, \citet{Qiu:2019} found evidence that the unfriendliness of maintainers in issue and PR comments is seen as a deterrent to newcomers.
% Compare to these studies, we found that the negative responses given to first-time contributors are majorly constructive and are not necessary a barrier to first-time contributors. 

\textbf{Developer Communication that impact contributions.}
% These related work is summarized into three different areas.
The first area is how discussions can  between developers.
The discussions of developers has been widely explored in GitHub (e.g., issue and PR) \citep{Bosu_2014,Tsay_2014}, issue tracking systems \citep{Correa_2013,Bertram_2010}, StackOverflow \citep{Barua_2014}, Twitter \citep{Bougie_2011}, and Reddit \citep{Iqbal_2021,Shrestha_2020}. 
As discussed by \citet{Tsay_2014}, the developer discussion can encourage further contributions to OSS projects. 
Developers are prone to encounter negative comments, which is not always lead to a detrimental outcome. 
Previous studies have focused on characterizing negative comments to investigate their impact.
\citet{Grigore_2011} analyzed what extent Wikipedia editor communications influence online trust among them.
They observed a significant difference in the amount of online trust among editors who share mainly positive or mainly negative sentiments.
\citet{ASRI:ist2019} investigated the role of sentiments
in code review activities and found that negative reviews took more time to be completed than positive and neutral reviews.
\citet{Miller:2021} conducted an empirical study on toxic GitHub issue discussions. 
In the same way, we also study about discussion between developer in GitHub, and  refers to this discussion as interaction in our work.
% In the same ways, our study investigate the correlation of negative responses toward future contribution.

% The third research area is related to toxic communication between developers.
Recent studies also reveal that developer communication can be entitled and demeaning complaints, arrogance.
At worst, these insults are common forms of toxicity, and these toxicity was not only written by people external to the projects, but project members were also common authors of toxicity.
\citet{Ferreira:2021} conducted a qualitative analysis on emails associated with reject changes from the linux kernel mailing list to understand confrontational conflicts in open source code review discussions.
Their results showed that incivility is common in code review discussions and uncivil comments can potentially be made by any people when discussing any topic.
\citet{Sanei:2021} investigated the impacts of sentiments and tones in GitHub issue discussions.
Similar to a study by \citet{Destefanis:2015}, they found evidence that positive comment sentiments were associated with shorter discussions and resolution time.
Although our work does perform some analysis of negative responses, we did not conduct a complete investigation on toxic responses.

\section{Conclusion}
\label{sec:conclusion}
The ability and sustain new contributors are crucial for an OSS's livelihood.  In this study, we investigate the correlation to characterize first responses as being positive; but delayed and that negative first responses are comment, suggestion, and question that is constructive or criticizing. 
Executing the protocols of a registered report, we are answered three research questions. 
Our results show that responses are mostly positive, with negative responses usually constructive criticism as feedback.
Based on our results, we recommend that OSS projects and first contributors should not be afraid of negative sentiments, instead view them as being constructive comments and suggestions to help with getting their contributions accepted into the project.

\section*{Declaration of Interest}

\textbf{Funding-} This work has been supported by JSPS KAKENHI Grant Number JP20H05706, JP20K19774\\

\noindent\textbf{Conflicts of interest-} Raula Gaikovina Kula is on the Editorial Board.\\

\noindent\textbf{Availability of replication package-}  The datasets generated during and/or analysed during the current study are available in the FirstResponsePR repository, https://github.com/NAIST-SE/FirstResponsePR. \\

\noindent\textbf{Author contributions-}
\begin{itemize}
    \item \textit{Noppadol Assavakamhaenghan}: Conceptualisation, Methodology, Investigation, Data collection, Qualitative Analysis, Original Writing draft, Visualisation.
    \item \textit{Supatsara Wattanakriengkrai}:  Investigation, Qualitative Analysis, Original Writing draft, Review.
    \item \textit{Naomichi
Shimada}: Investigation, Qualitative Analysis, Review.
    \item \textit{Raula Gaikovina Kula}: Conceptualisation, Funding Acquisition, review and editing drafts, Supervision, project administration.
    \item \textit{Takashi Ishio}:  Funding Acquisition, review and editing drafts, Supervision, project administration.
    \item \textit{Kenichi Matsumoto} : Funding Acquisition, review and editing drafts, Supervision, project administration.
\end{itemize}

% \section*{Acknowledgement}
% This work is supported by Japanese Society for the Promotion of Science (JSPS) KAKENHI Grant Numbers 20K19774 and 20H05706.
%\bibliographystyle{ieeetr}      % basic style, author-year citations
\bibliographystyle{spbasic}      % basic style, author-year citations
\bibliography{bibliography.bib}   % name your BibTeX data base

\end{document}